\documentclass[pre,preprint,aps,]{revtex4}
\usepackage{subcaption}
\usepackage{graphicx}
\usepackage{comment}
\usepackage{amsmath,amssymb,amsthm} 
\usepackage{bm}
\usepackage{verbatim}
\usepackage{float}

\begin{document}
\title{ The  stability  of galaxies in an expanding universe obtained by Newtonian dynamics}
\author{ S\o ren  Toxvaerd }
\affiliation{ Department
 of Science and Environment, Roskilde University, Postbox 260, DK-4000 Roskilde, Denmark}

\date{\today}
\noindent{\it published}: Clas. Quantum. Grav. 39, 225006 (2022), doi.org/10.1088/1361-6382/ac987f 
\vspace*{0.7cm}

\begin{abstract}
The dynamics of galaxies in an expanding universe is often determined for gravitational  and dark matter in an Einstein-de Sitter universe,
or alternatively by  modifying the gravitational long-range attractions in the Newtonian dynamics (MOND). Here the time evolution of galaxies
is determined by simulations of systems with pure gravitational forces by classical Molecular Dynamic simulations.
A time reversible algorithm  for  formation and aging of gravitational systems by self-assembly of baryonic objects,
recently derived  (Eur. Phys. J. Plus 2022, 137:99), is extended to include the Hubble expansion of the space. The algorithm is stable
for billions of time steps without any adjustments. The algorithm is used to simulate  simple  models of the Milky Way with the Hubble
expansion of the universe, and the galaxies are simulated for  times which corresponds to more than 25 Gyr. The rotating galaxies
lose bound objects from time to time, but they are still  stable at the end of the simulations.
The simulations indicate that the explanation for the dynamics of galaxies may be  that the universe is very young in cosmological times.
Although the models of the Milky Way are rather stable at 13-14 Gyr, which corresponds to the
cosmological time of the universe, the Hubble expansion will sooner or later  release the  objects in the galaxies. But
the simulations indicate that this will first happen in a far away future.	
\end{abstract}
\maketitle
\vspace{2pc}

\section{Introduction}

The constituents of the universe are often found in clusters of objects,  
  in planetary systems and in galaxies. The galaxies with stars  were
  established shortly after the ''Big Bang" \cite{Kolb2018}. 
 The two  groupings of objects have a common feature, the  objects rotate
 around their  center of gravity. But   a rotating galaxy
 seems, however, to be unstable unless it  contains a significantly larger
 mass (dark matter) than given by the known  baryonic matter in the galaxy   \cite{Martino2020}.

 The universe expands with a Hubble velocity \cite{Hubble1929}, and the expansion of the universe affects the
 dynamics of objects with gravitational forces.
 Simulations of galaxies with dark matter  in an expanding universe have been performed for many decades. In
 \cite{Klypin1983,Centrella1983} the authors solved the dynamics of baryonic objects  
 by the  Particle-Particle/Particle-Mess (PPPM)  method  \cite{Hockney1974}, 
   where each mass unit is treated as  moving in the collective field of all others, and the  Poison equation
 for the PPPM grid is solved numerically. The dynamics are obtained for  Zeldovich's adiabatic expansion of the Einstein-de Sitter  universe \cite{Zeldovich1970}.
 Later, simulations  with  large scaled computer packages with PPPM \cite{Diemand2008,Alimi2012} are with  many billions of  mass units.
  The evolution of galaxies has also been obtained from hydrodynamical large scaled cosmological simulations  \cite{Schaye2010,Dubois2014,Vogelsberger2014},  
and with 
co-evolving dark matter gas  and  stellar objects \cite{Dubois2016,Ludlow2021}. Cosmological simulations of galaxy formation is reviewed in \cite{Vogelsberger2019}. 

 An alternative theory for the dynamics of the galaxies, but without dark matter is the modified Newtonian dynamics (MOND),
 where Milgrom assumed that the  very weak gravitational force experienced by a star in the outer regions 
 of a galaxy varies inversely linearly with the distance, as opposed
 to the inverse square of the distance  in Newton's law of gravity \cite{Milgrom1983}.
 Merging  galaxies with MOND and with galaxies embedded in
 dark matter shows differences in  the merging time-scales, but only very small differences in
 the final stellar distribution \cite{Nipoti2007}. The  
 MOND simulations techniques for modified gravity is given in \cite{Llinares2018}. The MOND
 theory is extensively tested and compared with
 the Lambda cold  dark matter model $\rm{\Lambda}$CDM and in
 favour of the MOND theory  \cite{Kroupa2015,Banik2022}.

 Newtonian cosmology is derived in \cite{Ellis2014,Ellis2015}.
  Galactic dynamics with  pure Newtonian gravity forces has been compared with $N$-body MOND simulations, 
 and in  favour of  gravitational dynamics \cite{Wu2007,Derakhshani2014}, and 
Angus et al. find  similarly
 some shortcomings (mass-to-light ratios) of MOND at simulations of
the Carina Dwarf Spheroidal Galaxy orbiting the Milky Way
\cite{Angus2014}. 
The Newtonian dynamics simulations of galaxy systems described below  complement the existing analyzes of the stability of galaxies.
The  $N$-body simulations are for   models of celestial  objects in the Milky Way, and the objects interact
with pure gravitational forces and with and without the Hubble expansion of the space.

 The dynamics of models for galaxies are based on an  algorithm for 
 the classical dynamics of  systems of celestial objects  \cite{Toxvaerd2022}. Here (see Appendix A) the algorithm 
 is extended to include the Hubble expansion of the space. The discrete algorithm is time reversible and 
stable for billions of time steps without any adjustments.
The algorithm is used to simulate  simple  models of the Milky Way with and without the Hubble expansion of the universe.
The different models of galaxies are simulated for times which correspond to more than twice the age of the Universe and
both the galaxies without- as well as galaxies with the Hubble expansion of the space are stable. 
The simulations indicate that an alternative explanation to the dynamics of galaxies with dark matter is,
that the universe is very young in cosmological time units given by one rotation of the galaxy.
A galaxy  as old as our universe lose bound objects, 
but although the Hubble expansion  sooner or later will  release the stars in the Milky Way,
the simulations show that it will first happen in the distant future.

 \section{The  expanding universe} 
 A galaxy, $l$,  far away in the universe   moves away from  the Earth at a speed proportional to its ''proper  distance", $H r_{kl}(t)$,
to the Earth, $k$, measured at the  ''cosmological time" $t$, and this behaviour 
is explained by an expanding universe.
The Hubble constant $H$ is the  expansion coefficient in Hubble's law  \cite{Hubble1929,Lan2022}
\begin{equation}
  	\textbf{v}^{H}(\textbf{r}_{kl}(t))  =H \textbf{r}_{kl}(t)
\end{equation}
for the velocity of the increase of the distance $ \textbf{r}_{kl}(t)$ from the Earth  to a galaxy.
The Hubble expansion can be obtained by an intensive expansion of the space 
 independently of the baryonic matter in the universe
\begin{equation}
	 	\textbf{v}^{H}(\textbf{r}(t))=H \textbf{r}(t),
 \end{equation}
 by which the distance between pairs of positions $\textbf{r}_{k}(t),\textbf{r}_{l}(t),$ or $\textbf{r}_k(t),\textbf{r}_k(t+\delta t)$ increases with
 the Hubble velocity.  Eq. (2) fulfills   the \textit{Cosmological principle} and the 
 \textit{Copernican principle} for expansion of the universe \cite{Book}. The Cosmological principle, which was first formulated by Newton \cite{Newton1687}  demands,
 that no place in the universe is preferred,
 and the Copernican principle 
 demands, that no direction in the universe is preferred.  Eq. 2 will accelerate the expansion of the universe, and 
 observations of galaxies indicates, that the distances between the galaxies accelerate (dark energy)\cite{Riess1998,Perlmutter1999}.

The Hubble constant $H$ is  quoted in $\textrm{km s}^{-1}\textrm{Mpc}^{-1}$, for the velocity in $\textrm{km s}^{-1}$ 
of a galaxy 1 megaparsec ($3.09 \times 10^{19}$ km) away.
Its value is  \cite{Soltis2021}
\begin{equation}
	H=72.1 \pm  2.0 \  \textrm{km} \ \textrm{s}^{-1} \textrm{Mpc}^{-1}.
\end{equation}

For the Solar system with distances in au
(1 au$ \approx$ Earth's distance to the Sun =149597870.700 km (definition)) the Hubble constant is $33.93\times 10^{-11} $ $ \textrm{km s}^{-1}\textrm{au}^{-1}$,  and
the Earth's mean velocity  in the Solar system is $\approx 30$ $ \textrm{km s}^{-1}$,   so the Hubble velocity is about $\approx 10^{-11}$ of the Earth's velocity.
This small Hubble velocity has no effect on the orbits of the planets and the  stability of a planetary system (see  later).

The Milky Way is a barred spiral galaxy. The extension of the barred disk is 
2.5-3 kpc \cite{Rix2013},  and  the extension of the halos is  $\approx$ 100 000 pc, but some recent investigations
indicate that this   might be substantial larger, $\approx$ 300 kps or even more \cite{Deason2020,Li2021}.
The  Milky Way rotates with a rotational
velocity $ 220 \pm 10$  $ \textrm{km s}^{-1}$ for stars at a distance of  $ 8.0 \pm 0.3 10$  kpc from its center  \cite{Camarillo2018},
but the rotations at different distances from the
center of the galaxy deviate from calculations of 
 the rotation caused   by the  gravitational forces between its constituents, and this deviation is explained by the existence of 
 unknown dark matter \cite{Weinberg1984,Chiba2020,Martino2020}.
 The Milky Way contains very old stars from the period just after the ''Big  Bang" \cite{Bond2013}, so the galaxy has presumably exist
 for  more than thirteen billion years. The rotational time (galactic year) of the Milky Way is
 estimated to be $\approx$ 220-240 million years, and this implies
 that an old  star in the Milky Way with this rotational velocity   must have performed of the order $\approx $ 60 cycles.

  The Hubble velocity between pairs of stars in the Milky Way
  at distances of e.g. $5\times 10^{-2}$ Mpc  is 3.5 $ \textrm{km s}^{-1}$, and it is  not negligible compared with  the rotational velocity
  of the galaxy. The expansion of the
  universe is with a Hubble velocity, which is of the order $\approx 1 \% $ or more  of the rotational velocity of the Milky Way,
  and the Hubble expansion will affect the rotation and the stability of the galaxy.

\begin{figure} 
	\begin{center}	
	\includegraphics[width=12cm,angle=-90]{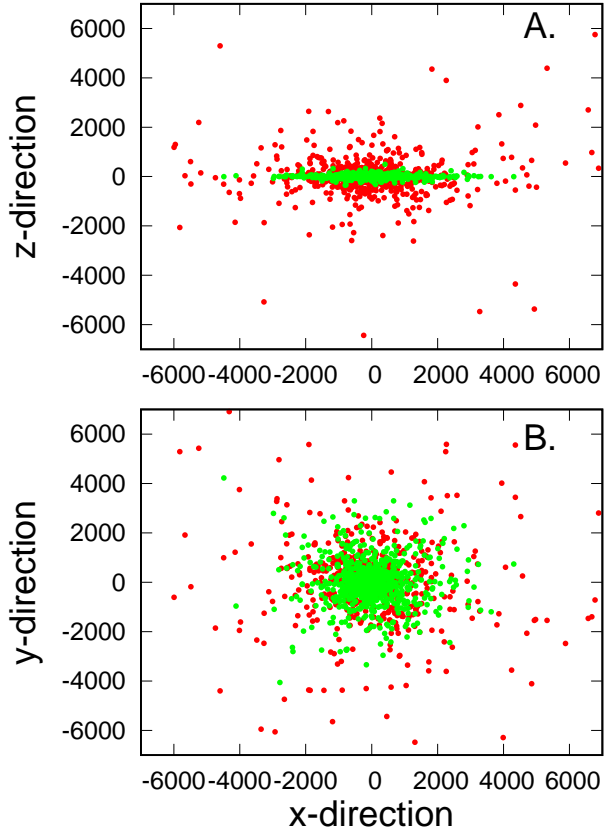}
		\caption{ Positions of the objects. A: Site view of the system; B: Top view. The green dots are the positions
			at the start and the red dots are the positions at $t=2.5\times 10^4 (10^7$  time steps).}
	\end{center}			
\end{figure} 
\section{Simulation of  galaxies  in an expanding universe}

A galaxy and the Milky Way contains hundred of billion of stars, and a substantial amount of baryonic 
gas \cite{Gupta2012,Bergma2018} and
it is not possible  directly to obtain the Newtonian dynamics of a galaxy   with this number of  objects. Instead, we have simulate
  models of small ''galaxies" of hundred of objects in
   orbits around their center of gravity, and in an expanding space with various strength of the expansion.
A recent article describes how to create  
  a collection of objects with rotation about their center of gravity  by  Molecular Dynamics (MD) simulations, obtained by an extension of 
  Newton's discrete algorithm \cite{Toxvaerd2022}. The algorithm   is further developed to also include the Hubble expansion of space,
   and the algorithm with
  the extensions are given in the Appendix. 
	      	     
The collection  of  baryonic objects with Newtonian dynamics will, for many starting positions and velocities of the objects 
spontaneously form a system with many of the  objects in rotations  about a heavy mass in its center. If these MD systems   shall simulate  the   dynamics of  a galaxy in the expanding  universe, then one must relate $\textit{distances, times,}$  and $\textit{Hubble expansion}$ in the MD systems with
	     the corresponding distances, times and Hubble expansion in a galaxy.

\subsection{Determination of distances, times and  $H$ for the Hubble expansion in the MD systems}
 A system $without$ $a$ $Hubble$ $expansion$ of the space is
  obtained from ensembles of  $N=1000$  objects  with a disk-shaped  distribution of
   their start positions and    with $\approx$ zero velocities in the z-direction. 
    The distribution of the objects at the start and after the creation of
     a system with $\approx 300$ of the objects in  circulations around a heavy object  at the center of mass
      is shown  in Figure 1. 
\begin{figure}
\begin{center}	
\includegraphics[width=6cm,angle=-90]{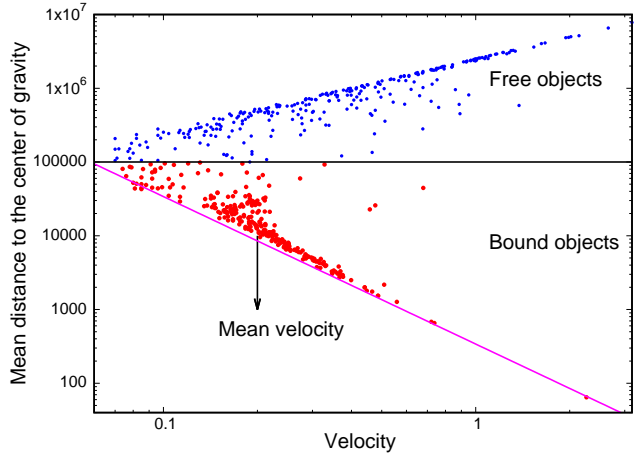}
\caption{ $log$ mean distances  to the  mass center    as a function of the
$log$ mean velocities  of the objects.  The bound objects (red spheres)
are located on the lower branch of the distribution and the upper branch shows (blue spheres) the mean locations of the free objects.
Magenta line: $log(y)=log(G m_c/x^2)$.}
\end{center}		
\end{figure}

The system with positions shown in Figure 1 is simulated for a long period of time. The system is evolving and with  mergers  
   of objects, and after 
 $t=2.5\times 10^6$ it contains  $N=557$ objects, and with  270 of them in
 circulations  around the 
 heavy mass in the center.
 The mean distances, $\bar{r}_{i,c}$, in the succeeding time interval $\Delta t \in[2.5 \times 10^6, 3.0 \times 10^6]$,  as a function
 of the mean velocity, $\bar{v}_{i,c}$, relative to the center $c$ is shown in Figure 2.
 A bound 
 object will have a relative 
 small mean distance  
 to the center of mass in contrast to a free object, which has a large and monotonic increasing distance to  
  the center. The lower
 branch of mean distances in Figure 2 are for the  collection of bound objects, and the the upper branch are the mean distances of the free objects moving
 away from the center  with constant velocities. An object, $i$, in an elliptic orbit around
 a heavy gravitational center with
 the mass $m_c$ will have a mean velocity
$\bar{v}_{i,c} = \sqrt{G m_c/a_i}$,
 where $a_i$
 is the semimajor axis in its ellipse. The magenta line is
 the line $log(y)=log( Gm_c/x^2)$ with the mass $m_c=340$ of the  center of the system (for  units in the MD systems see the Appendix).
 The systematic deviation of the mean distances of the bound objects
 from the straight line for increasing distances to the center is partly
 caused by, that the eccentricities of the  orbits of the bound objects increase with increasing distance to the center of mass, but 
 the distribution of objects for greater distances $2000 < \bar{r}_{i,c}$  shows,
 that the objects are not lined up in a  row
 of objects in elliptical orbits  as in a  planetary system.

 The mean distances, rotation times  and velocities for the bound objects,
 shown in Figure 2 are used to relate the MD system with the Milky Way.

The Milky Way has an extension of $\approx 100000 $ pc, but the extension of some of the halos may be more. The lower branch in Figure 2 with the bound
objects is within a MD distance $100000$ MD length units or less from the center. So a  relation between  the
distances in the Milky Way in pc and the distances in the MD system is  1 ps= 1 MD length unit.

The galactic year of the Milky Way  is about 220 million years \cite{Camarillo2018}, and
the Milky Way contains a very old star \cite{Bond2013}, which has been created shortly after the ''Big Bang".
A star with a rotation time of 220 million years, must  have performed   only  $\approx$ 60 cycles in the Milky Way after it is created. 
 The distance  $R_0$ of our planet system to the center of the Milky way is $R_0=8.34 \pm 0.16$ kpc \cite{Reid2014}.
 These data are used to relate the cosmological time with the time in the MD systems.
    The rotation time of  objects in the MD system at $\approx$ 8000 MD length units
    is $t_{orbit} \approx 1.3 \times 10^5$, so  these objects have performed $\approx 60$ rotations
    at the MD time $8.0 \times 10^6$. The  Figure 2   is for 0.3125 of the cosmological time of the universe, or
$\approx$ 4.2 Gyr, and the age of the universe corresponds to $\approx$  $8.0 \times 10^6$ in the MD systems.

\begin{figure}
\begin{center}	
\includegraphics[width=6cm,angle=-90]{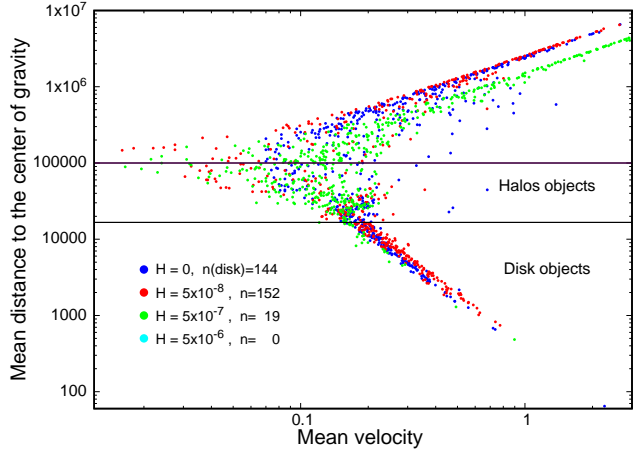}
\caption{The $log$ mean distances $log(\bar r_{i,c})$ of the  objects as
a function of their $log$ mean velocities  $log(\bar v_{i,c}) $ with respect to the center $c$  
 for different values of the Hubble constant $H$. The mean positions are obtained for the time
interval $\Delta t \in  [2.5 \times 10^6, 3.0 \times 10^6]$.}
\end{center}	
\end{figure}

The mean velocity $\bar{v}$ of the bound objects in the MD system in Figure 2 is $\bar{v} \approx 0.15$. 
The value  of the Hubble expansion constant in the MD system is obtained from the
ratio   $H/\bar{v}$ between the Hubble constant per unit length and the
rotational velocity  of the galaxy as

\begin{equation}
	(H/ \bar{v})_{\textrm{MD}}=(H/ \bar{v})_{\textrm{Milky Way}},
\end{equation}
which determines the value of $H$ in the MD system,
\begin{eqnarray}
	H= 0.15 \times 72 \times 10^{-6} \textrm{km}\ \textrm{s}^{-1}\ \textrm{pc}^{-1}/220\ \textrm{km}\ \textrm{s}^{-1} \nonumber \\
		\approx 5.\times 10^{-8} \textrm{ per MD length unit}.
\end{eqnarray}
Let the estimates   of the extent of the Milky Way, the extent of the  MD model of a galaxy,
the rotation time of the Milky Way and the mean velocity in the MD system all be with
an uncertainty with a factor  of two, then the total uncertainty of the estimate of $H$ in the MD systems is $\approx 2^4 \approx 10 $. The value of the
Hubble constant in the MD units   for the model of  the  Milky Way in the expanding universe  is 
\begin{equation}
	H= 5.\times 10^{-8\pm 1}.
\end{equation}

\begin{figure}
	\begin{center}	
\includegraphics[width=6cm,angle=-90]{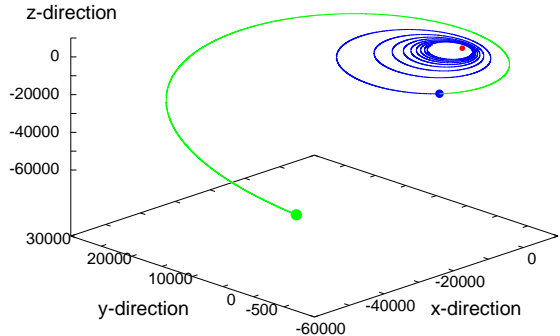}
\caption{The release of the last bound object $l$ from the
	system with $H=5. \times 10^{-7} $. Blue: The orbit for $\bar r_{l,c}(t) < 15000$ and green:
	for $\bar r_{l,c}(t) > 15000$. The center of gravity with red. The blue sphere is the position at $t=6.5 \times 10^6$, and the green sphere is the position at $t=7.75 \times 10^6$.}
	\end{center}		
\end{figure} 

\subsection{The formation and stability of  galaxies obtained by classical dynamics}

The MD system without an expansion of the space, with positions  shown in Figure 1 at  $t=0$ (green) and $t = 2.5 \times 10^4$ (red),   and Figure 2 at $t = 2.5 \times 10^6$ is  formed spontaneously. 
 This system  is used to investigate the formation and stability
 of MD systems to various values of $H$ in the Hubble expansion. Systems with $H= 0,$ $ 5.
 \times 10^{-8},$ $ 5.  \times 10^{-7}$ and $5.  \times 10^{-6}$,
  respectively,  are simulated and their behavior is shown below.

 In Figure 3  the distributions of objects  at $t=2.5 \times 10^6$ is shown for various strength of the Hubble constant.
 The four systems are  started with
 the  disk formed  distribution of 1000 objects with equal masses, shown with green dots in Figure 1. 
 All four  systems spontaneously formed gravitational systems. The systems contain  many bound objects with different masses and
 in circulation around their centre of gravity, but the system with
 $H = 5. \times 10^{-6}$ was unstable and with no bound  objects already after $ t=1.0 \times 10^6$.
 The radius of the  disk in the Milky way is $\approx 15$ kpc \cite{Rix2013} corresponding to a  a distance 15000 MD length units in the 
 MD systems. At $ t=2.5 \times 10^6$, the number of objects
 within this distance to the center of the galaxy are  144, 152, 19 and 0, respectively for the four different values of the expansion.
 The system with $ H= 5.  \times 10^{-7}$  and only 19 objects with  $ \bar r_{i,c} <15000$ is, however,
 also unstable  and without any objects with distances less than 15000, but first after $t=6.5 \times 10^6$ (Figure 4).

\begin{figure}
\begin{center}	
\includegraphics[width=6cm,angle=-90]{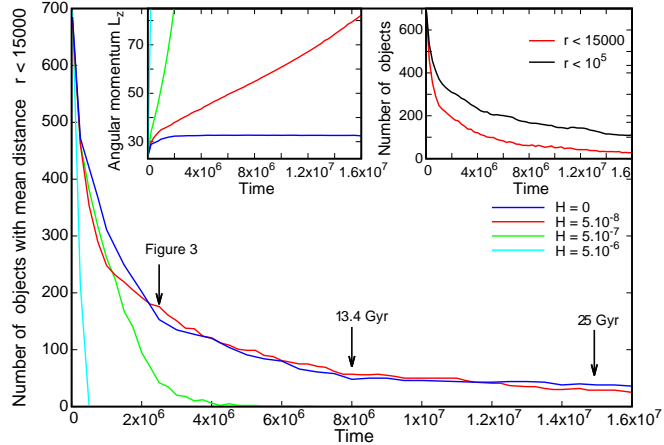}
\caption{ Number of  objects $i$ with mean distances $\bar r_{i,c}(t) < 15000$ to the center of the galaxy $c$ as a function of  time. The systems
are  started from the ''gas" distribution shown with green dots in Figure 1.
The colors (same as in Figure 3) are for  different strength of of the Hubble expansion. The inset to the left shows the time evolution of the $L_z$-components of
  the  angular momenta  of the galaxies  with  different strength  of the Hubble expansion.
	In the right inset the number of bound objects with  $\bar r_{i,c}(t) < 15000$  for $H=5.\times 10^{-8}$ are compared with the total number of bound objects
	 $\bar r_{i,c}(t) < 10^5$ in the galaxy.}
\end{center}
\end{figure}

 Figure 5 shows the time evolution of the number of objects within a mean distance $\bar{r}_{i,c} < 15000$ from the center  $c$. 
Almost all of the objects  are within this distance
at the start
 of the evolution, but as the self-assembly and aging of the Newtonian systems evolves,
 the number  declines slowly  for $ H \le 5. \times 10^{-8}$, whereas it
 rapidly goes to zero for faster expansions. The arrow to the left
 marks the time $2.5 \times 10^{6}$ with the distribution of objects shown in Figure 3, and the arrow in the middle is at the time   $8.0 \times 10^{6}$, which 
 should correspond to the age of our universe. The systems with $H \le 5. \times 10^{-8}$
 are rather stable at this time, but  with a weak declining number of objects
 within the  distance  $\bar{r}_{i,c} < 15000$.
 The systems are started from the same start distribution, shown with green spheres in Figure 1, and the self-assembly with
 merging of objects results in slightly different masses of the center of the galaxies and small differences between the numbers
 for $H=0$ and $H=5. \times 10^{-8}$.
The left inset  in Figure 5 shows the
 $L_z$ components of the angular momenta $\textbf{L}$ for the galaxies with  the different strength of  the  expansion.
 The angular momentum of a system without a Hubble expansion is  conserved
  for Newton's discrete dynamics (blue curve in the inset), but the Hubble expansion increases the  angular  momenta,  and
   this will sooner or later destroy the galaxies (Figure 4).   The number    of bound object with mean distance $\bar r_{i,c}(t) < 15000$ to the center of the galaxy 
     with $H=5. \times 10^{-8}$ is compared with the corresponding total number  of all bound objects ( objects with  $\bar r_{i,c}(t) < 10^5$ )
     in the galaxy in the  right inset  of Figure 5.
    The two sets of numbers decreases with time, but very slowly   and the destruction of the system with first happen in a far away future for the
     galaxy with  $H=5. \times 10^{-8}$.

 A series of control simulations have been performed in order to
 ensure, that the results shown in Figure 5 are representative for the formation and aging of  gravitational systems with Hubble expansion.
 In one set of  test simulations the three different expansions was first included in the
 dynamics at $t= 2.5 \times 10^6$ in the system without expansion.   But
 with the same result that the system with  $ H = 5. \times 10^{-8}$ was rather stable, but the two systems with stronger Hubble expansions were unstable.
 In another set of simulations the Hubble expansion
 was added to a Newtonian system with another distribution of the objects at the start of the
 formation of the galaxy, but with the same result: All the systems are simulated for times much  longer, that
 corresponds to ours cosmological time  $\approx $13.4 Gyr,
 and the systems are still rather  stable   for an expansion with
 $H \le 5. \times 10^{-8}$,  which should correspond to the Hubble expansion of the universe. 
\begin{figure}
	\begin{center}	
\includegraphics[width=6cm,angle=-90]{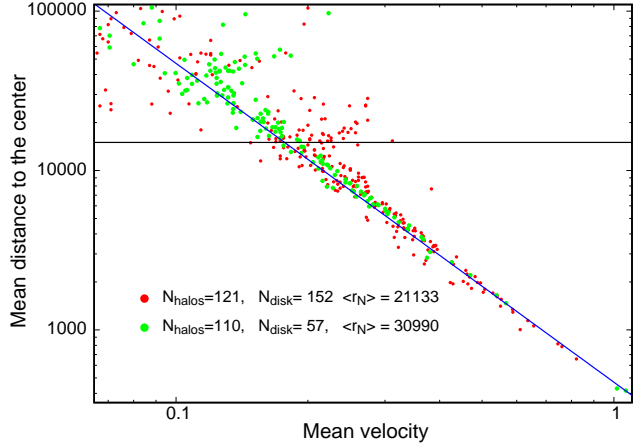}
\caption{   $log(\bar r_{i,c})$  as
a function of   $log(\bar v_{i,c}) $ 
and at different times. With red is for
$t=2.5 \times 10^6$, and with green is  for $t=8. \times 10^6$ ($\approx$ 13.4 Gyr).
$\textrm{N}_{\textrm{disk}}$ are for $ \bar r_{i,c} <15000$ and  $\textrm{N}_{\textrm{halos}}$ are for $ 15000 < \bar r_{i,c} <100000$.
$<\bar r_{i,c}>$ is the mean positions for the   $\textrm{N}_{\textrm{disk}}
+\textrm{N}_{\textrm{halos}}$ objects. The blue line,  $r=m_c/v^2$, is the line for  mean positions of objects in ellipses, and with the same eccentricities.}
	\end{center}
 \end{figure} 

 Figure 6 shows the $log$ mean distances $log(\bar r_{i,c})$ of the  objects as
 	a function of their $log$ mean velocities  $log(\bar v_{i,c}) $ with respect to the center $c$  of the
	system with $H=5.\times 10^{-8}$, and at different times. With red is for
$t=2.5 \times 10^6$, also shown in Figure 3, and with green is  for $t=8. \times 10^6$ ($\approx$ 13.4 Gyr).
The numbers $\textrm{N}_{\textrm{disk}}$ are for objects with $ \bar r_{i,c} <15000$ and the numbers $\textrm{N}_{\textrm{halos}}$ are for 
 objects with $ 15000 < \bar r_{i,c} <100000$. $<\bar r_{i,c}>$ is the mean distances for the   $\textrm{N}_{\textrm{disk}}
+\textrm{N}_{\textrm{halos}}$ objects.
 The blue line,
$log(r)=log(G m_c/v^2)$
is the line for  mean distances of objects in ellipses with the same eccentricities. The distributions of the objects deviates from 
this line  for a ''Kepler-like" order with the objects in unperturbed elliptical orbits around the center of mass.
In particular, the objects with large distances to the center of gravity deviate from the straight line
 by  having in general a higher velocity   than given by the blue line, and the aging of the system does not remove this tendency. The 
system at $t= 8. \times 10^6$ should correspond to the cosmological time 13.4 Gyr of the universe, and the systems are still aging with
losses of bound objects and with a monotonic increasing rotational velocities, as shown in the inset of Figure 5 for $L_z(t)$.

\section{Discussion}

The dynamics of galaxies in an expanding universe are often determined for gravitational  and dark matter in an Einstein-de Sitter universe,
or alternatively by  the modified Newtonian dynamics (MOND)\cite{Milgrom1983}. But here the dynamics of galaxies is obtained for
systems with pure gravitational matter by classical Molecular Dynamic simulations.
The  dynamics of the systems is not explained by dark matter, nor by a modification of the gravitational
attractions at long distances, but by that the universe is a young universe measured in cosmological times.
The Milky Way is believed to be  more than 13 Gyr,
but it corresponds to only $\approx$ 60 rotations or galactic years, and it is
nothing compared to the age of our solar system with 4.56 Gyr, i.e. number of
rotations of the Earth. The stability of the solar system is still 
to debate \cite{Batygin2008,Mogavero2021},
 and in the light of these facts the universe must be characterized as a very young universe.
 The dynamics and aging of the MD systems with gravitational matter shows, however, that the objects in the small  MD systems are not in a stable steady
 state after only sixty rotations, so how can hundred of billion of stars and a substantial amount of baryonic gas then be it?

 According to the present investigation,
 the explanation for the dynamics of the universe could be, that
the galaxies  are formed spontaneously at the relatively high 
concentration of baryonic matter shortly after the Big Bang. The universe expands and
the galaxies are  aging, but  the  galaxies are not in  a steady state at a time which corresponds to 13.4 Gyr (Figure 5, Figure 6). 
The systems with- and without   a Hubble expansion 
still occasionally lose   bound objects at  13.4 Gyr, and the mean distance of the bound objects increase also over time. 
The bound objects far away from the center of gravity are not distributed  as in a stable
Newtonian system. They have in general a higher velocity, than if they were ordered in simple Kepler-like orbits (Figure 6).
The hypothesis can perhaps be tested by statistical analyses of the density and velocity distribution (distribution of light intensity) in galaxies as  
functions of the age of the galaxies.

 The Hubble expansion increases the angular momentum of a galaxy and this fact implies, that the galaxy sooner
 or later will be destroyed, and that all the objects will be free in an open universe. The end of the
 system with $H=5. \times 10^{-7}$ is shown in Figure 4, with the release of the last bound object in the system. But this system has a strength of
 the Hubble expansion, which probably is ten times stronger than in the universe.
 The simulations  indicate, however, that for the systems  with  a Hubble expansion similar to the Hubble expansion of the universe, this destruction
 might first happen in a far away future. These systems are 
 simulated for  what corresponds to more than 25 Gyr, and they still contain many
 objects with $\bar r < 15000$ (Figure 5).

The algorithm is well suited for
analyzes of formation and aging of  gravitational systems in an expanding universe.
The central difference algorithm, Eqn. (A11)+ (A3), is time reversible,
and the underlying Newtonian discrete algorithm, which is used in almost all
MD simulations in physics and chemistry is exact in the sense, that it has  the same
 invariances as Newton's analytic dynamics \cite{Toxvaerd2022,Toxvaerd2014}.
 The algorithm is stable, the present simulations are with more than $6 \times 10^9$ discrete time steps
 for each systems, and each simulation has taken 4-5 months, but without any adjustments.

\appendix

\section{A time reversible   algorithm for discrete classical dynamics in  an expanding space} 
\subsection{Newton's discrete central difference algorithm}\label{sec: discrete dynamics }
 In Newton's classical discrete dynamics \cite{Newton1687,Toxvaerd2020}  a new  position $\textbf{r}_k(t+\delta t)$ at time $t+\delta t$ of an object
$k$ with the mass $m_k$  is determined by
the force $\textbf{f}_k(t)$ acting on the object   at the discrete positions $\textbf{r}_k(t)$  at time $t$, and  
 the position $\textbf{r}_k(t-\delta t)$ at $t - \delta t$  as
\begin{equation}
	 m_k\frac{\textbf{r}_k(t+\delta t)-\textbf{r}_k(t)}{\delta t}
			=m_k\frac{\textbf{r}_k(t)-\textbf{r}_k(t-\delta t)}{\delta t} +\delta t \textbf{f}_k(t),	
 \end{equation}
where the momenta $ \textbf{p}_k(t+\delta t/2) =  m_k (\textbf{r}_k(t+\delta t)-\textbf{r}_k(t))/\delta t$ and
 $  \textbf{p}_k(t-\delta t/2)=  m_k(\textbf{r}_k(t)-\textbf{r}_k(t-\delta t))/\delta t$ are constant in
the time intervals in between the discrete positions.
Newton postulated Eq. (A1) and obtained his  second law, and the analytic dynamics  in the  limit $ lim_{\delta t \rightarrow 0}$.

 The algorithm, Eq. (A1), is usual presented  as the ''Leap frog" algorithm for the velocities
\begin{equation}
\textbf{v}_k(t+\delta t/2)=  \textbf{v}_k(t-\delta t/2)+ \delta t/m_k  \textbf{f}_k(t).
\end{equation}
The positions
are determined from the discrete values of the momenta/velocities as
\begin{equation}
\textbf{r}_k(t+\delta t)= \textbf{r}_k(t)+ \delta t \textbf{v}_k(t+\delta t/2).	  
\end{equation}	  

Newton's discrete algorithm, Eq. (A1), has been re-derived many times and with different names, and almost all Molecular Dynamics (MD) simulations
in physics and chemistry as well as many simulations in astrophysics
are performed with his discrete algorithm.
The rearrangement of Eq. (A1) gives e.g. the ''Verlet" algorithm \cite{Verlet1967}.

  The classical discrete dynamics between $N$ spherically symmetrical objects
     with masses $ m^N=m_1, m_2,..,m_N$ and positions $\textbf{r}^N(t)=\textbf{r}_1(t), \textbf{r}_2(t),..,\textbf{r}_N(t)$  is obtained 
 by Eq. (A2) and Eq. (A3). The momentum and angular momentum for a conservative system of the $N$ objects is conserved due to Newton's third law, and
the energy is  also conserved \cite{Lee1983,Toxvaerd1994,Toxvaerd2014}, so Newton's time reversible and symplectic  discrete dynamics has the same invariances as his analytic dynamics.
Newton's discrete algorithm make it possible to simulate systems of interacting  objects for billions of time steps without $any$ adjustments for the infinite
range of the gravitational forces or corrections for symplecticity,
time reversibility,  conservation of momentum, angular momentum and energy.

\subsection{Newton's discrete central difference algorithm with merging of gravitational objects}
The discrete central difference algorithm for merging of objects is derived in \cite{Toxvaerd2022}. 
Let all the spherically symmetrical objects
 have the same (reduced)  number density $\rho= (\pi/6)^{-1} $ by which
  the diameter $\sigma_i$ of the spherical object $i$ is 

   \begin{equation}
	   	 		 \sigma_i= m_i^{1/3}
				  \end{equation}
				    and   the collision diameter 
				    \begin{equation}
					    	\sigma_{ij}=	\frac{\sigma_{i}+\sigma_{j}}{2}.
				    \end{equation}	
				     If  the distance $r_{ij}(t)$ at time $t$ between two objects is less than $\sigma_{ij}$ 
				     the two objects merge to one spherical symmetrical object with mass

				     \begin{equation}
					     m_{\alpha}= m_i + m_j,
				     \end{equation}	 
				     and diameter
				     \begin{equation}
					      \sigma_{\alpha}= (m_{\alpha})^{1/3},
				     \end{equation}
				     and with the new object $\alpha$  at the position
				     \begin{equation}
					     \textbf{r}_{\alpha}(t)= \frac{m_i}{m_{\alpha}}\textbf{r}_i(t)+\frac{m_j}{m_{\alpha}}\textbf{r}_j(t),
				     \end{equation}	
				     at the center of mass of the the two objects before the fusion.
				     (The   object $\alpha$ at the center of mass of the two merged objects $i$ and $j$ might occasionally be near another object $k$
				     by which more objects merge, but after the same laws.)

The momenta  of the objects in the discrete dynamics just before the fusion are $\textbf{p}^N(t-\delta t/2)$ and the
total momentum of the system is conserved  at the fusion if
\begin{equation}
	\textbf{v}_{\alpha}(t-\delta t/2)= \frac{m_i}{m_{\alpha}}\textbf{v}_i(t-\delta t/2)+ \frac{m_j}{m_{\alpha}}\textbf{v}_j(t-\delta t/2),
\end{equation}
which determines the  velocity $\textbf{v}_{\alpha}(t-\delta t/2)$ of the merged object.

The algorithm for the system of  baryonic objects  consists of the equations (A2)+(A3) for time steps without merging of objects, and the fusion of objects
is given by the equations (A5),(A6), (A7), (A8) and (A9).
The discrete algorithm with merging of objects  has the same invariances as Newton's exact analytic and his discrete dynamics \cite{Toxvaerd2022}.

\subsection{The algorithm for discrete classical mechanics in the expanding universe}
Newton's discrete dynamics
changes the position  of an object $k$.
If the space expands monotonously over time with
the Hubble velocity $\textbf{v}^H$, then the expansion also changes   the distance between two positions.  The new position  $\textbf{r}_k(t+\delta t)$ is the
sum of the change due to the gravitational force on $k$ and the contribution from the Hubble expansion.
The mean location of an object   changes  from
$\textbf{r}_k(t-\delta t/2)=(\textbf{r}_k(t-\delta t)+\textbf{r}_k(t))/2$ at $t \in[t-\delta t, t]$ to
$\textbf{r}_k(t+\delta t/2)=(\textbf{r}_k(t)+\textbf{r}_k(t+\delta t))/2$ at  $t \in [t, t+ \delta t]$  .
The Hubble expansion changes the  distance between  the  two  positions by
the Hubble velocity
\begin{eqnarray}
\textbf{v}^H= H \delta \textbf{r}_k = H(\textbf{r}_k(t+\delta t/2)-\textbf{r}_k(t-\delta t/2)) \nonumber \\
=\delta tH \frac{\textbf{r}_k(t+\delta t)-\textbf{r}_k(t)}{2\delta t}+
\delta tH \frac{\textbf{r}_k(t)-\textbf{r}_k(t-\delta t)}{2\delta t} \nonumber  \\
=\frac{\delta t H}{2} \textbf{v}_k(t+\delta t/2) + \frac{\delta t H}{2} \textbf{v}_k(t-\delta t/2).
\end{eqnarray}
By including the Hubble velocity, Eq. (A10),  in the
Newtons algorithm, Eq. (A2), and after a re-arrangement,
one obtains the algorithm for the classical mechanics with a Hubble expansion of the space included in the Newtonian dynamics
\begin{equation}
\textbf{v}_k(t+\delta t/2)=\frac{(1+ \delta t H /2)\textbf{v}_k(t-\delta t/2) + \delta t/m_k\textbf{f}_k(t)}{1-\delta t H/2}.
\end{equation}
The discrete classical dynamics with Hubble expansion, Eqn. (A11)+(A3), is still time reversible, but it  increases
the velocities, the momenta and the angular momenta.

   The small  galaxies in the articles are obtained for thousand objects,
  which at the start of the simulation, at $t=0$ are separated with a mean distance $<r_{ij}(0)> \approx 1000$ and with a Maxwell-Boltzmann
  distributed velocities with  mean velocity $ <\mid \textbf{v}_i(0) \mid > \approx 1$.
  The gravitational strengths in the article  are in units of the gravitational constant $G=1$ and the mass unit $m=m_i(0)=1$ and diameters
    of the objects $\sigma_i(0)=1$.
  For the  set-up of the systems see also \cite{Toxvaerd2022}. 
  The systems are followed more than $6.4 \times 10^9$ MD time steps, i.e. $t=1.6 \times 10^7$ time-units, which
   corresponds to 25 Gyr  of our universe.
\\
$   $ \\
$\textbf{Acknowledgements}$
This work was supported by the VILLUM Foundation Matter project, grant No. 16515.
Isaac Newton is gratefully acknowledged.
\\
$   $ \\
$\textbf{Data Availability Statement}$ Data will be available on request.
\\
$  $ \\
$\textbf{References}$
\\

\end{document}